\documentclass[a4paper,fleqn,usenatbib]{mnras}
\usepackage[T1]{fontenc}
\usepackage{ae,aecompl}
\usepackage{amsmath}
\usepackage{amsfonts}
\usepackage{amssymb}
\usepackage{graphicx}
\usepackage{color,xcolor}
\def\mnras{MNRAS}
\def\aap{A$\&$A}
\def\nat{Nature}
\def\apj{ApJ}
\def\apjl{ApJ Letters}
\def\apjs{ApJS}

\def\aj{AJ}
\def\icarus{Icarus}

\newcommand{\me}{\, {\rm M}_{\oplus}}
\newcommand{\msun}{\, {\rm M}_{\odot}}
\newcommand{\msunyr}{\, {\rm M}_{\odot}\,{\rm yr^{-1}}}
\newcommand{\rsun}{\, {\rm R}_{\odot}}
\newcommand{\au}{\, {\rm au}}
\newcommand{\ab}{\, a_{\rm bin}}

\newcommand{\acav}{\, {r_{\rm c}}}
\newcommand{\ecav}{\, {e_{\rm c}}}
\newcommand{\rcavapo}{\, {r_{\rm c,a}}}

\title[How to form TOI-1338/BEBOP-1?]{Constraining the formation history of the TOI-1338/BEBOP-1 circumbinary planetary system}
\author[G. A. L. Coleman et al]{Gavin A. L. Coleman$^1$\thanks{Email: gavin.coleman@qmul.ac.uk},
Richard. P. Nelson$^1$,
Amaury H. M. J. Triaud$^2$,
\newauthor Matthew R. Standing$^{3,4}$\\
1. Astronomy Unit, Department of Physics and Astronomy, Queen Mary University of London, Mile End Road, London, E1 4NS, UK\\
2. School of Physics and Astronomy, University of Birmingham, Edgbaston, Birmingham B15 2TT, UK\\
3. School of Physical Sciences, The Open University, Milton Keynes, MK7 6AA, UK\\
4. European Space Agency, ESAC, Spain}
\date{Accepted 2023 October 17. Received 2023 October 16; in original form 2023 August 4}
\pubyear{2023}
\begin{document}
\label{firstpage}
\pagerange{\pageref{firstpage}--\pageref{lastpage}}
\maketitle
\begin{abstract}
The recent discovery of multiple planets in the circumbinary system TOI-1338/BEBOP-1 raises questions about how such a system formed.
The formation of the system was briefly explored in the discovery paper, but only to answer the question do current pebble accretion models have the potential to explain the origin of the system?
We use a global model of circumbinary planet formation that utilises N-body simulations, including prescriptions for planet migration, gas and pebble accretion, and interactions with a circumbinary disc, to explore the disc parameters that could have led to the formation of the TOI-1338/BEBOP-1 system.
With the disc lifetime being the main factor in determining how planets form, we limit our parameter space to those that determine the disc lifetime. These are: the strength of turbulence in the disc, the initial disc mass, and the strength of the external radiation field that launches photoevaporative winds.
When comparing the simulated systems to TOI-1338/BEBOP-1, we find that only discs with low levels of turbulence are able to produce similar systems.
The radiation environment has a large effect on the types of planetary systems that form, whilst the initial disc mass only has limited impact since the majority of planetary growth occurs early in the disc lifetime.
With the most TOI-1338/BEBOP-1 like systems all occupying similar regions of parameter space, our study shows that observed circumbinary planetary systems can potentially constrain the  properties of planet forming discs.

\end{abstract}

\section{Introduction}
\label{sec:intro}

With space telescopes such as {\it Kepler} \citep{Borucki10} and {\it TESS} \citep{TESS-Paper} and dedicated observation programmes such as {\it BEBOP} \citep{Martin19}, the number of confirmed circumbinary planets, those orbiting a pair of stars, has dramatically increased over the last decade (e.g. Kepler-16b, -34b, -35b and TOI-1338b to name a few)  \citep{Doyle11,Welsh12,Kostov20}.
The most recent such planet discovered is BEBOP-1c, a 65$\me$ planet on a 215.5 d orbit, found in a multiple planetary system with the less massive TOI-1338b \citep{Standing23}.
Interestingly, circumbinary planets currently being discovered are found to be orbiting close to the zone of dynamical instability \citep{Holman99,Langford23}, a limit in a circumbinary system, within which lies an unstable region where orbiting objects would be excited on to eccentric orbits, leading to collisions with the central star, or ejection from the system.
Additionally, in those systems where multiple planets have been discovered, the innermost planets have been found to be orbiting near the zone of dynamical instability \citep{Martin18}.
Such a pile-up of planets at the edge of the zone of dynamical instability appears to not be due to observational bias with the transit method, and instead hints at possible origins in how the planets form and evolve in these systems \citep{Martin14}.
Attempts to form such planets include the {\it in situ} model and one that involves concurrent growth and migration from further out in the system.

The {\it in situ} scenario suffers from a number of issues that likely hinder the formation of planets near to the instability zone, including: gravitational interactions with non-axisymmetric features within circumbinary discs leading to large impact velocities between planetesimals \citep{Marzari08,Kley10}; differential pericentre alignment of eccentric planetesimals of different sizes that leads to corrosive collisions \citep{Scholl07}; excitation of planetesimal eccentricities through N-body interactions resulting in large relative velocities, which are disruptive for accretion onto planetary bodies \citep{Meschiari12a,Meschiari12b,Paardekooper12,Lines14,Bromley15}.
Ways to overcome the problems with {\it in situ} formation have been explored, including having extremely massive protoplanetary discs \citep{Marzari00,Martin13,Meschiari14,Rafikov15}, or if the fragments are reaccreted and form second or later generations of planetesimals \citep{Paardekooper10}.
More recently, it has been shown that {\it in situ} pebble accretion scenarios also suffer from difficulties because a parametric instability can generate hydrodynamical turbulence that stirs up pebbles, rendering pebble accretion onto planetary embryos inefficient \citep{Pierens20,Pierens21}. 

In forming the planets at large orbital distances and allowing them to migrate to their current locations, numerous works have shown that migrating planets in circumbinary discs stall when they reach the central cavity, with the precise stopping location depending on parameters such as the planet mass \citep{Nelson03,Pierens07,Pierens08a,Pierens08b,Thun18,Penzlin21}.
Recently, \citet{Coleman23} presented a global N-body planet formation model that includes prescriptions for pebble accretion, planet migration, gas accretion and evolution of a circumbinary disc.
They demonstrated that the model could form planets similar to Kepler-16b and Kepler-34b.
Indeed in \citet{Standing23}, the authors used the model of \citet{Coleman23} to show that the cores of the observed planets in the TOI-1338 system could be formed through pebble accretion in the outer disc, far from their observed locations, before the planets simultaneously accreted gas and migrated towards the cavity region around the central stars.
Their main aim was to form systems similar to that observed, and so only coarsely examined a range of parameters, these being the initial disc mass and the metallicity.

In this work, we expand on the parameters chosen in \citet{Standing23}, by examining more broadly, what parameters can lead to the formation of systems such as TOI-1338/BEBOP-1.
The aim of this approach is not only to determine what combinations of parameters are capable of forming such systems, but also those that are not, hence providing hints about the properties of planet forming discs that can be compared to observations.
We follow \citet{Coleman23} and use the N-body code {\textsc mercury6} \citep{Chambers,ChambersBinary} that is coupled to a viscously evolving 1D disc model, along with prescriptions for photoevaporation, pebble accretion, planet migration, gas accretion, and effects arising from interactions with a central binary.
Our models are able to produce systems similar to TOI-1338/BEBOP-1, with the best-fitting systems forming in discs with low levels of turbulence, and in weaker radiation environments.

This paper is laid out as follows.
We outline our physical model in Sect \ref{sec:base_model}, whilst we describe our population parameters in Sect. \ref{sec:pop_parameters}.
In Sect. \ref{sec:pop_results}, we outline the results of our population models.
Finally, we discuss our results and draw conclusions in Sect. \ref{sec:conclusions}.

\section{Physical Model}
\label{sec:base_model}

In the following section, we provide a basic overview of the physical model we adopt and the numerical scheme used to undertake the simulations.
The N-body simulations presented here were performed using the {\sc{mercury6}} symplectic N-body integrator \citep{Chambers}, updated to accurately model planetary orbits around a pair of binary stars \citep{ChambersBinary}.
We utilise the `close-binary' algorithm described in \citet{ChambersBinary} that calculates the temporal evolution of the positions and velocities of each body in the simulations with respect to the centre of mass of the binary stars, subject to gravitational perturbations from both stars and other large bodies.
We also include prescriptions for the evolution of 1D protoplanetary disc as well as disc-planet interactions.
With the disc model being 1D in nature, we also include prescriptions that take into account non-axisymmetric effects (i.e. a precessing eccentric inner disc cavity) due to the binary stars.
The full model and the additional prescriptions due to the binary can be found in \citet{Coleman23}, with Table. \ref{tab:sim_param} showing the applicable parameters, but we briefly describe the model below:

\begin{table}
\centering
\begin{tabular}{l|lc}
Parameter & Description & Value\\
\hline
$M_{\rm A}\ (\msun)$ & Primary Mass & $1.0378^1$\\
$M_{\rm B}\ (\msun)$ & Secondary Mass & $0.2974^1$\\
$T_{\rm A}\ ({\rm K})$ & Primary Temperature & 4300\\
$T_{\rm B}\ ({\rm K})$ & Secondary Temperature & 3300\\
$R_{\rm A}\ (\rsun)$ & Primary Radius & 2\\
$R_{\rm B}\ (\rsun)$ & Secondary Radius & 1.5\\
$a_{\rm bin}\ (\au)$ & Binary Separation & $0.129^1$\\
$e_{\rm bin}$ & Binary Eccentricity & $0.156^1$\\
Metallicity (dex) & Stellar Metallicity & 0\\
& & \\
$\acav\ (\ab)$ & Cavity Radius & 3.7377\\
$\ecav$ & Cavity Eccentricity & 0.4162\\
$\rcavapo\ (\ab)$ & Cavity Apocentre & 5.2933\\
$C_1$ & Cavity Parameter 1 & 1.1\\
$C_2$ & Cavity Parameter 2 & 0.32\\
$C_3$ & Cavity Parameter 3 & 4.5\\
R$_{\rm in}\ (\au)$ & Disc Inner Boundary & 0.129\\
R$_{\rm out}\ (\au)$ & Disc Outer Boundary & 200\\
$f_{41}$ & Ionising EUV photon Parameter & 100\\
$r_{\rm g} (\au)$ & Gravitational Radius & 11.81\\
\end{tabular}
\caption{Simulation parameters. $^1$\citet{Kostov20}}
\label{tab:sim_param}
\end{table}

(i) We solve the standard diffusion equation for a 1D viscous $\alpha$-disc model \citep{Shak,Lynden-BellPringle1974}.
Disc temperatures are calculated by balancing black-body cooling against viscous heating and stellar irradiation from both stars.
The viscous parameter $\alpha_{\rm visc}=10^{-3}$ throughout most of the disc, but increases close to the central stars to mimic the eccentric cavity that is carved out by the tidal forces of the central stars.
When giant planets are present, tidal torques from the planets are applied to the disc leading to the opening of gaps \citep{LinPapaloizou86}.

(ii) We incorporate models of photoevaporative winds removing material from the disc, both internally driven through radiation emanating from the central stars, and externally driven due to FUV radiation from nearby sources (i.e. O/B--stars).
For internal photoevaporation due to EUV radiation from the central stars, we include a standard photoevaporative wind model \citep{Dullemond}, where the wind is assumed to be launched thermally from the disc upper and lower surfaces beyond a critical radius that approximately corresponds to the thermal velocity being equal to the escape velocity from the system.
With ionising radiation not just impacting the disc from the central stars, but also from nearby stars in the local star-forming region \citep[e.g.][]{Haworth18,Haworth23}, we include external photoevaporation in the models to account for the effects of ionising FUV photons.
We adopt the model found in \citet{Matsuyama03}, which drives a wind outside of the gravitational radius where the sound speed in the heated layer is $T\sim$1000~K.

(iii) The main source of the accretion of solids in our models is through pebble accretion. We follow the pebble accretion model of \citet{Johansen17}, where a pebble production front moves outwards in the disc over time. This production front arises from dust particles coagulating and settling to the disc midplane forming pebbles. Once these pebbles become large enough, they begin to drift inwards through gas drag forces, thus creating a pebble production front when the drift time-scale is equal to the growth time-scale.
As the pebbles drift inwards they can be accreted by planetary embryos, allowing them to grow on short time-scales \citep{Lambrechts12}.
The accretion of pebbles continues until the planets reach the pebble isolation mass, that being the mass where planets are able to sufficiently perturb the local disc, forming a pressure bump exterior to the planet's orbit, that traps pebbles and halts pebble accretion on to the planet \citep{Lambrechts14,Ataiee18,Bitsch18}.

(iv) Eccentric cavities, arising because of the tidal torque from the central binary, have been seen in observations and numerical simulations of circumbinary discs \citep[e.g.][]{Artymowicz94,Dutrey94,Pierens13,Mutter17D,Thun17,Coleman22b}.
The shape and size of these cavities depends on the binary properties, i.e. mass ratio and eccentricity, and local disc properties such as the viscosity parameter $\alpha$ \citep{Kley19}.
The main effect of circumbinary disc cavities on planet formation is the creation of a planet migration trap as the corotation torque is increased due to the positive surface density gradient at the cavity edge.
To simulate the effects of an eccentric cavity in our 1D disc model, we ran 2D hydrodynamical simulations of circumbinary discs using \textsc{fargo3d} \citep{FARGO-3D-2016} to determine the cavity structure. In the 1D models, we simulate the azimuthally averaged surface density profile of the cavity by adjusting the viscosity parameter $\alpha$, whilst maintaining a constant gas flow rate through disc. This forms an inner cavity in the disc and leads to a buildup of material at the outer edge of the cavity, as required.

(v) Using the above 2D hydrodynamical simulations, we take into account the precessing, eccentric nature of the inner disc cavity in 1D models through construction of 2D maps of the gravitational acceleration experienced by test particles embedded in the disc due to the non-axisymmetric density distribution. We also create maps of the gas surface densities and gas velocities.
These maps are used when integrating the equations of motion of planets and when calculating relative velocities between planets and drifting pebbles. These effects are mainly relevant near the cavity region.

(vi) We use the torque formulae from \citet{pdk10,pdk11} to simulate type I migration due to Lindblad and corotation torques acting on planetary embryos.
Corotation torques arise from both entropy and vortensity gradients in the disc, and the possible saturation of these torques is included in the simulations.
The influence of eccentricity and inclination on the migration torques, and of eccentricity and inclination damping are included \citep{Fendyke,cressnels}.

(vii) Type II migration of gap forming planets is simulated using the impulse approximation of \citet{LinPapaloizou86}, where we use the gap opening criterion of \citet{Crida} to determine when to switch between type I and II migration.
Thus, when a planet is in the gap opening regime, the planet exerts tidal torques on the disc to open a gap, and the disc back-reacts onto the planet to drive type II migration in a self-consistent manner.

(viii) The accretion of gaseous envelopes on to solid cores
occurs once a planet’s mass exceeds 1$\me$.
We utilise the formulae based in \citet{Poon21} that are based on fits to gas accretion rates obtained using a 1D envelope structure model \citep{Pap-Terquem-envelopes,PapNelson2005,CPN17}.
To calculate these fits \citet{Poon21} performed numerous simulations, embedding planets with initial core masses between 2--15 $\me$ at orbital radii spanning 0.2--50 $\au$, within gas discs of different masses.
This allowed for the effects of varying local disc properties to be taken into account when calculating fits to gas accretion rates, a significant improvement on fits used in our previous work \citep[e.g.][]{ColemanNelson14,ColemanNelson16,ColemanNelson16b}.
We use these fits until a planet is massive enough to undergo runaway gas accretion and open a gap in the disc.
The gas accretion rate is then limited to either the maximum value of the fits from \citet{Poon21}, or the viscous supply rate.
All gas that is accreted onto a planet is removed from the surrounding disc, such that the accretion scheme conserves mass.

\section{Population Parameters}
\label{sec:pop_parameters}

Whilst the work of \citet{Standing23} only varied the initial disc mass and metallicity parameter, we alter and broaden our parameter study here.
Given that \citet{Standing23} showed that discs with Solar metallicity are capable of forming planetary systems similar to TOI-1338/BEBOP-1, and given that the observed metallicity of the system is expected to be around Solar \citep{Kostov20,Standing23}, we only consider discs of Solar metallicity in this study.
With the disc lifetime being one of the main considerations when forming planets, since it controls the length of time over which a planet can both accrete and migrate, we mainly concentrate on parameters affecting this property \citep{Winter22,Qiao23}.
Namely, this involves exploring different values for initial disc masses, varying mass loss rates due to external photoevaporation, as well as different values of the viscosity parameter $\alpha$.
For these parameters we take random values between the limits shown in table \ref{tab:pop_param}, with the last column denoting whether we randomise in log or linear space.

We choose initial disc masses between 5 and 15\% of the combined binary mass, with the maximum disc mass being equal to the most massive disc a star can host before it becomes gravitationally unstable \citep{Haworth20}.
Numerous works have provided observational estimates for the viscosity parameter $\alpha$ \citep{Isella09,Andrews10,Pinte16,Flaherty17,Trapman20,Villenave20,Villenave22}.
We adopt values of $\alpha$ that are consistent with such estimates \citep[see][for a recent review]{Rosotti23}.
For the final parameter that affects the disc lifetime, the rate of external photoevaporation, we use mass loss rates for a 100 $\au$ disc of between $10^{-9}$--$10^{-6}\msunyr$.
These values correspond to FUV field strengths of between $\sim30$--$\sim3\times10^4$, consistent with what is expected across star forming regions such as Orion, with other low mass regions such as Taurus and Lupus occupying the lower  region of our parameter space \citep{Winter18}.

\begin{table}
\centering
\begin{tabular}{l|ccc}
Parameter & Lower Value & Upper Value & Dimension\\
\hline
$M_{\rm disc}\ (M_{\rm bin})$ & 0.05 & 0.15 & linear\\
$\dot{M}_{\rm pe,ext}\ (\msunyr)$ & $10^{-9}$ & $10^{-6}$ & log\\
$\alpha_{\rm b}$ & $10^{-4}$ & $10^{-2.5}$ & log\\
\end{tabular}
\caption{Values for the parameters varied amongst the populations.}
\label{tab:pop_param}
\end{table}

We include 42 planetary embryos in the simulations, spaced equidistantly between 2--20 $\au$.
Initially the planetary embryos are assigned masses according to the transition mass where planets begin to accrete from their entire Hill sphere instead of just the Bondi sphere:
\begin{equation}
    M_{\rm trans} \sim \eta^3 M_{\rm bin}
\end{equation}
where $\eta$ is the dimensionless measure of gas pressure support \citep{Nakagawa86}, and $M_{\rm bin}$ is the combined binary mass.
To initialise the planetary embryos we assume that they are equal to 0.1$\times M_{\rm trans}$, since recent work has shown that the largest embryos that form through gravitational collapse of pebble clouds in protoplanetary discs are generally 10-30 times smaller than the transition mass \citep{Schafer17,Abod19,Liu20,Coleman21}.
For the planetary eccentricities and inclinations we initialise them randomly between 0--0.001, and 0--0.36$^{\rm o}$, respectively.
We initialise the surface density in the circumbinary discs using $\Sigma_{\rm g}(r) = \Sigma_{\rm g,1\au}(r/1\au)^{-1}$, where $\Sigma_{\rm g,1\au}$ is the surface density at $1\au$ assuming the disc has a size up to 200$\au$ and an initial $M_{\rm disc}$.
We run each simulation for 10 Myr to account for the entire protoplanetary disc phase, and additionally allowing for dynamical evolution of the systems after the dispersal of the circumbinary discs.

\begin{figure*}
\centering
\includegraphics[scale=0.5]{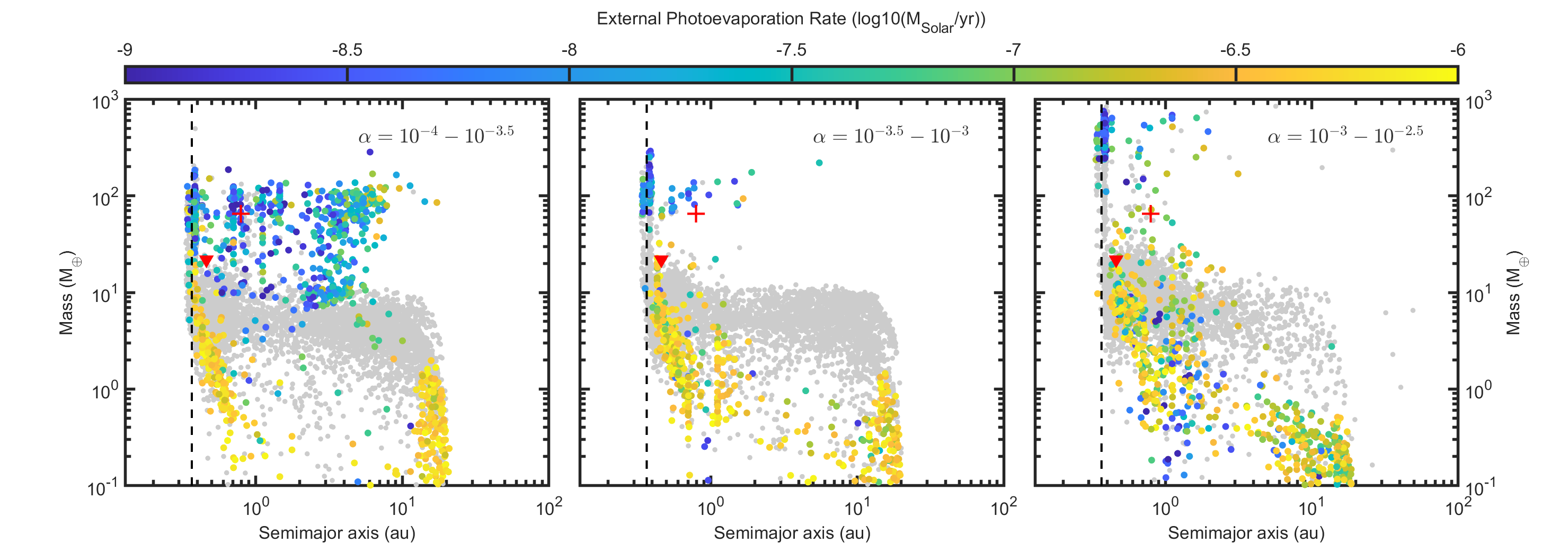}
\caption{Mass versus semimajor axis plot for all planets formed in simulations with viscosity $\alpha$ values between: $10^{-4}$--$10^{-3.5}$ (left panel), $10^{-3.5}$--$10^{-3}$ (middle panel), and $10^{-3}$--$10^{-2.5}$ (right panel). The colour coding denotes the external photoevaporation rate experience by the planet forming disc, whilst grey points show planets that have been lost from the simulations, either through collisions or ejections. The red symbols denote the mass and semimajor axis of TOI-1338b and BEBOP-1c. The dashed black line denotes the stability limit \citep{Holman99}.}
\label{fig:mva_alpha}
\end{figure*}

\section{Results}
\label{sec:pop_results}

The main objective of this work is to use the TOI-1338/BEBOP-1 planetary system to explore the disc parameters that give rise to its formation.
Whilst we do not discuss the formation and evolution of the circumbinary systems in detail, instead just focusing on the final outcomes, the formation pathways and behaviours are generally similar to those found in \citet{Coleman23}. There, planetary embryos accrete pebbles and migrate towards the central cavity, becoming trapped there due to strong corotation torques. This then allows collisions to occur, whilst providing time for giant planet cores to accrete gas and become gas giants.
Below we begin by analysing our simulation results as a whole population before using the TOI-1338/BEBOP-1 planetary system to determine which parameter choices are conducive to its formation.

\subsection{Exploring the parameter space}

\subsubsection{Effects of changing $\alpha$}
In fig. \ref{fig:mva_alpha} we show the final masses versus semimajor axes for all planets that formed in our simulations. The left-hand panel shows the planets from simulations with $\alpha \le 10^{-3.5}$, the middle panel is for $10^{-3.5}<\alpha\le10^{-3}$ and the right-hand panel is for $\alpha > 10^{-3}$.
Planets denoted by grey points show those lost from the simulations, either through collisions or ejections following interactions with the central binary.
The colour coding shows the external photoevaporation rate of the disc that each specific planet formed in.
Finally the red triangle and plus sign indicate the locations of TOI-1338b and BEBOP-1c, respectively.
The dashed black line denotes the outer edge of the zone of dynamical instability \citep{Holman99}.

As can be seen in fig. \ref{fig:mva_alpha}, there is a significant difference in the populations of planets that form in circumbinary discs, depending on the value of $\alpha$.
Looking at the left-hand panel of fig. \ref{fig:mva_alpha} representing low $\alpha$ values, it is clear that a large number of planets with masses of 10--200 $\me$ were able to form and migrate in close to the central binary.
Some of these planets were able to open gaps in the disc before they reached the cavity, and were then able to migrate outwards slightly as the outer disc photoevaporated, leaving only an inner disc with which the planets could exchange angular momentum.
This resulted in the large number of planets with masses $m_{\rm p}\sim 100\me$ orbiting between 2--6 $\au$.
A large number of planets with masses 10--200$\me$ are also found to be orbiting near the zone of dynamical instability (dashed line) and around where the inner cavities were located ($<1\au$).
These planets formed and migrated deeper into the cavity region, becoming trapped in that region due to positive corotation torques that arise from the positive surface density profiles associated with the cavities.
They then opened gaps in the disc and remained close to the binary as the inner disc accreted on to the binary.

When comparing the left-hand panel of fig. \ref{fig:mva_alpha} to the middle and right-hand panels, it is clear that increasing the viscosity affects the distribution of planets that form, both in terms of their masses and semimajor axes.
A noticeable difference is the number of giant planets that form in discs with higher $\alpha$ values (right-hand panel) and attain masses greater than 1 Jupiter mass.
This is due to these planets being able to undergo runaway gas accretion before they open deep gaps that inhibit the accretion of gas on to the planets.
For planets forming in lower $\alpha$ discs, a lower mass planet opens gap in the disc, hindering the accretion of massive envelopes.
The semimajor axis distribution of massive planets is also seen to depend on $\alpha$.
As mentioned above, in the discs with low $\alpha$ values, some of the giant planets and planets with masses just below 100$\me$ were able to migrate outwards after dispersal through evaporation of the disc lying exterior to their orbits, leaving only the inner disc with which to exchange angular momentum, causing the planets to migrate outwards until the inner disc accreted on to the central stars.
This led to a larger dispersion in giant planet semimajor axes.

For the giant planets that form in the discs with higher $\alpha$ values, shown in the right panel of fig. \ref{fig:mva_alpha}, they are mainly found around the zone of dynamical instability or where the cavity was located, $\leq 1\au$.
Those giant planets with semimajor axes $>1\au$, are found in systems that contained multiple giant planets, and so dynamical interactions with inner giant planets halted their migration at a larger separation.
In the discs with an intermediate $\alpha$, the middle panel of fig. \ref{fig:mva_alpha}, there is a smooth transition in both the mass and semimajor axis distribution of giant planets.

\begin{figure}
\centering
\includegraphics[scale=0.6]{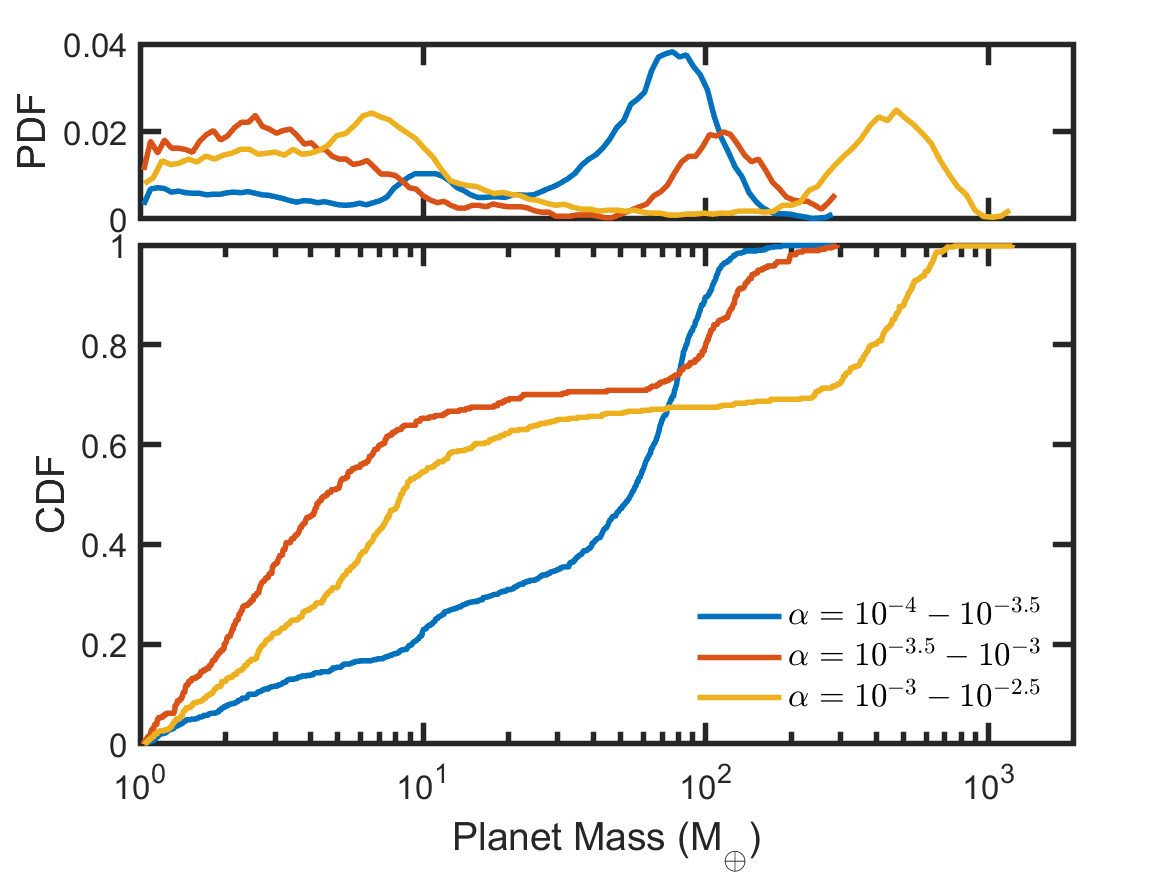}
\caption{Cumulative distribution functions (bottom panel) and probability distribution functions (top panel) of planet masses for those forming in discs with different ranges of the viscosity parameter $\alpha$. We show discs with low $\alpha$ values (blue line), a high $\alpha$ values (yellow line) and an intermediate value for $\alpha$ (red line). We only calculate the distributions for planets with masses greater than 1$\me$.}
\label{fig:planet_mass_dist}
\end{figure}

We now examine the change in populations for super-Earth and Neptune mass planets.
The differences in the distributions of these planets is harder to distinguish from fig. \ref{fig:mva_alpha}.
Therefore in fig. \ref{fig:planet_mass_dist} we show the cumulative distribution function (bottom panel) and the probability density function (top panel) of planet masses as a function of the viscous $\alpha$ value.
The blue line represents planets forming in discs with low values of $\alpha$ (equivalent to the left-hand panel in fig. \ref{fig:mva_alpha}), whilst the red and yellow lines show the intermediate and higher $\alpha$ discs, equivalent to the middle and right-hand panels of fig. \ref{fig:mva_alpha}, respectively.
The differences in distributions for giant planets as a function of the disc viscosity is clear in fig. \ref{fig:planet_mass_dist}, where only a small fraction (10.5\%) of the planets shown by the blue line have masses $m_{\rm p}>100\me$, whilst a larger fraction giants formed in the more viscous discs ($\sim20\%$ and $\sim32\%$ for the intermediate and high $\alpha$ values, respectively).

Moving on to super-Earths and Neptune mass planets, the fraction of super-Earths forming ($m_{\rm p}\le 15\me$) increases as $\alpha$ increases from its lowest to its intermediate value, and then decreases as the value of $\alpha$ is increased to its highest value. 
In the most viscous discs, more planets are able to undergo runaway gas accretion, and N-body interactions between multiple giant planets can lead to ejections, such that both of these effects reduce the fraction of super-Earths. 
This fraction rises from 28.5\% in discs with low $\alpha$ values to 67.5\% in discs with intermediate $\alpha$ values, before then dropping to 59.6\% for high $\alpha$ discs.
Interestingly, in high $\alpha$ discs, there appears to be a bi-modal distribution when combining the fractions of super-Earths with the fraction of giant planets, equating to 92\% of all planets.
This is especially evident in the top panel of fig. \ref{fig:planet_mass_dist} where the yellow line shows two distinct populations of planets, one centred around $\sim$Jupiter mass, and the other around $\sim7\me$.
The bimodality in planet mass is not as evident in the intermediate $\alpha$ discs but nonetheless can be seen (corresponding to $\sim 80\%$), and is not evident in the low $\alpha$ discs as there is a larger abundance of planets forming with masses 25--100 $\me$.
As discussed previously, the lack of giant planets in the lowest $\alpha$ discs is due to the planets opening gaps before they can undergo runaway gas accretion.

As well as the distributions for planet masses varying, there is also a difference when it comes to the distributions of semimajor axes for super-Earths forming in circumbinary discs with different values of $\alpha$.
From fig.~\ref{fig:mva_alpha} it is clear that as $\alpha$ increases, the spatial distribution of super-Earths begins to shift to planets orbiting closer to the zone of dynamical instability and the cavity edge.
Indeed $\sim96\%$ of super-Earths forming in discs with $\alpha>10^{-3.5}$ orbit with semimajor axes $<2\au$, whilst only 57.5\% orbit there in low $\alpha$ discs.
This change in orbital locations arises because more massive planets in the low $\alpha$ discs open gaps before they are able to undergo runaway gas accretion.
These planets then migrate toward the central stars slowly, or even outwards if the outer disc is photoevaporated, and act as a barrier for less massive super-Earths migrating inwards whilst remaining embedded in the circumbinary disc.
This allows the super-Earths to retain larger semimajor axes for longer, and ultimately remain there once the discs fully disperse.

\begin{figure*}
\centering
\includegraphics[scale=0.6]{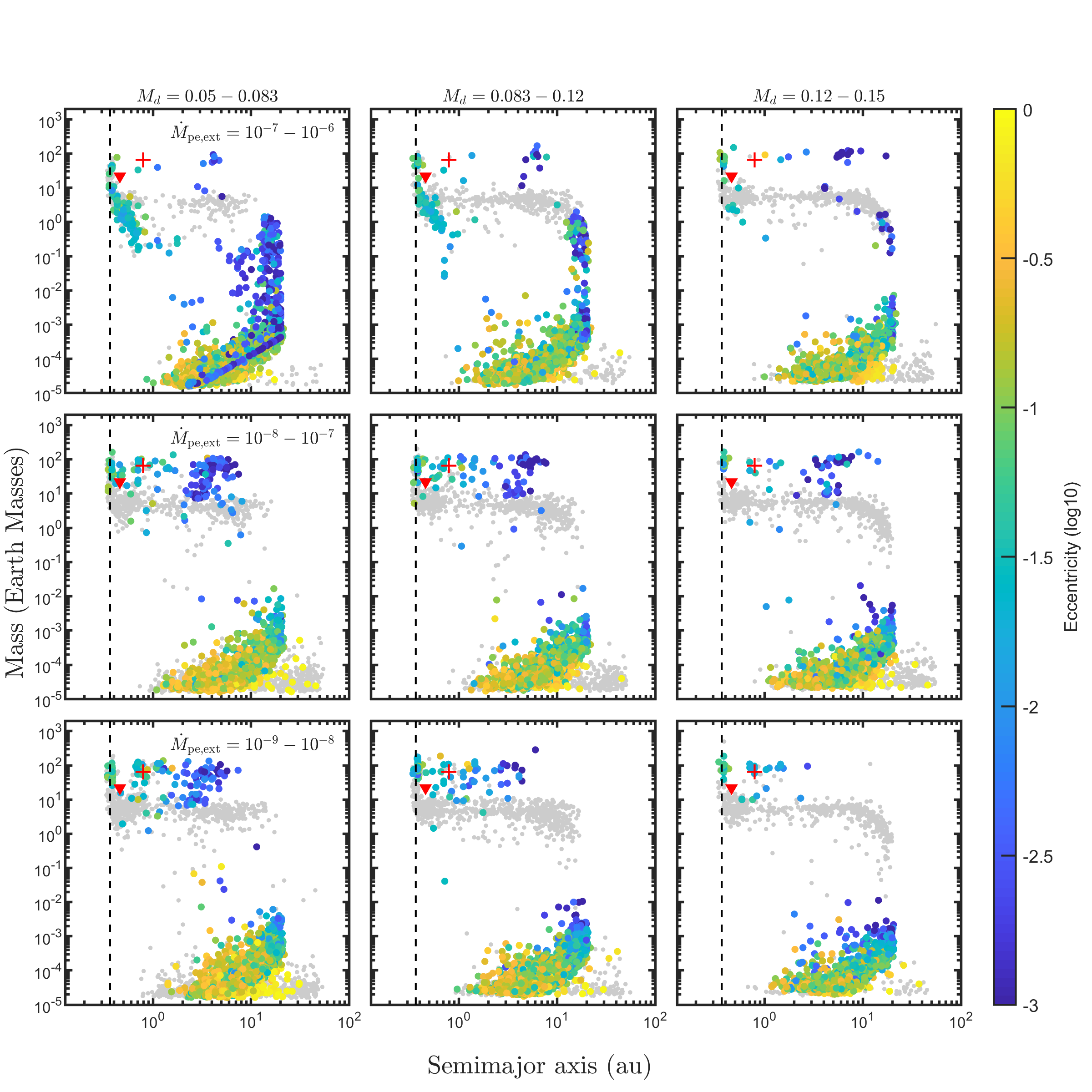}
\caption{Mass versus semimajor axis plots for all planets in simulations with specific parameters for the discs with viscosity $\alpha$ values of between $10^{-4}$--$10^{-3.5}$. Different panels show different initial disc masses in units of binary masses (columns) and external photoevaporation rates in units of $\msunyr$ (rows) with values for each row and column in the first row and column respectively. The colour coding denotes the final eccentricities of the planets, whilst grey points show planets that have been lost from the simulations, either through collisions or ejections. The red symbols denote the mass and semimajor axis of TOI-1338b and BEBOP-1c. The dashed black line denotes the outer edge of the zone of dynamical instability \citep{Holman99}.}
\label{fig:mva_large_4}
\end{figure*}

\subsubsection{Effects of initial disc mass and external photoevaporation}

Whilst the panels in fig. \ref{fig:mva_alpha} showed mass versus semimajor axis for discs separated by the $\alpha$ parameter, they did not show the differences that arise due to the initial disc mass or the external photoevaporation rate.
Both of these parameters, as well as $\alpha$, can determine the lifetimes of circumbinary discs, especially that of the outer regions, and so are important in determining the final mass and locations of planets that form within them.
In fig. \ref{fig:mva_large_4} we show the mass versus semimajor axis for all planets that form in low $\alpha$ discs  ($10^{-4}$--$10^{-3.5}$), with each panel representing a range of initial disc masses and external photoevaporation rates.
For example the top left panel shows the planets from simulations with initial disc masses between 0.05-0.083$\times M_{\rm bin}$ and external photoevaporation rates of $10^{-7}$--$10^{-6}\msunyr$.
The colour coding here represents the instantaneous eccentricity of the planets after 10 Myr.
Similar to fig. \ref{fig:mva_alpha}, the grey points show planets that have been lost through collisions or ejections, the dashed line shows the outer edge of the zone of dynamical instability, and the red symbols denote the mass and semimajor axis of TOI-1338b and BEBOP-1c, respectively.
Looking at the top left panel of fig. \ref{fig:mva_large_4} showing the planets forming in the least massive discs and in most intense external radiation environments, it is clear to see that few giant planets with masses greater than 100$\me$ were able to form.
The majority of planets that formed remained at sub-terrestrial masses and underwent minimal migration, hence the large number of planets between 10--20$\au$.
For planets that were able to grow, the majority of those were able to migrate in towards the central cavity, orbiting there once the disc fully dispersed.

When moving to the right across the top row of fig. \ref{fig:mva_large_4}, the initial disc mass increases.
As the initial disc mass increases, the amount of material available for accretion subsequently increases, which leads to more massive planets being able to form.
The disc lifetime is also increased which allows further time for planets to accrete gas and migrate.
In the top right panel of fig. \ref{fig:mva_large_4} a larger number of Neptune to Saturn mass planets were able to form and can be seen orbiting near the cavity, or around $\sim$few $\au$ after migrating outwards once they opened gaps in the disc.
A number of super-Earth and terrestrial mass planets were able to form and migrate to the cavity region, whilst a larger number of planets were ejected from the system (shown by the grey points) as N-body interactions increased in magnitude due to the more massive planets.

The effects of an increase in disc lifetime as the external photoevaporation rate decreases is even more evident when moving down the rows in fig. \ref{fig:mva_large_4}.
With external photoevaporation not dispersing the outer disc as quickly, more dust is available to be converted into pebbles.
This allows planets to accrete pebbles for longer, letting them attain more massive cores.
The increase in disc lifetime, again also gives more time for planets to accrete gas and migrate in the disc.
Looking at the bottom of fig. \ref{fig:mva_large_4}, it is clear that a large number of planets were able to accrete substantial gaseous envelopes and reach masses similar to Saturn, after they opened gaps in their discs.
Interestingly the panel with the largest fraction of planets in this mass regime is the bottom left, that showing the smaller disc masses and the weakest external photoevaporation rates.
Those planets are also found to be orbiting with a larger variation in orbital separation than those found in more massive discs, showing the influence that outward migration has had on these planets.
Outward migration was more of a factor here, since the discs were initially less massive, and so the outer regions dissipated more quickly than did the more massive discs, allowing planets to migrate outwards and not interact as strongly as those found in more massive discs, where inward migration was prominent for longer.
This ultimately led to more massive planets surviving in the systems.

\begin{figure*}
\centering
\includegraphics[scale=0.6]{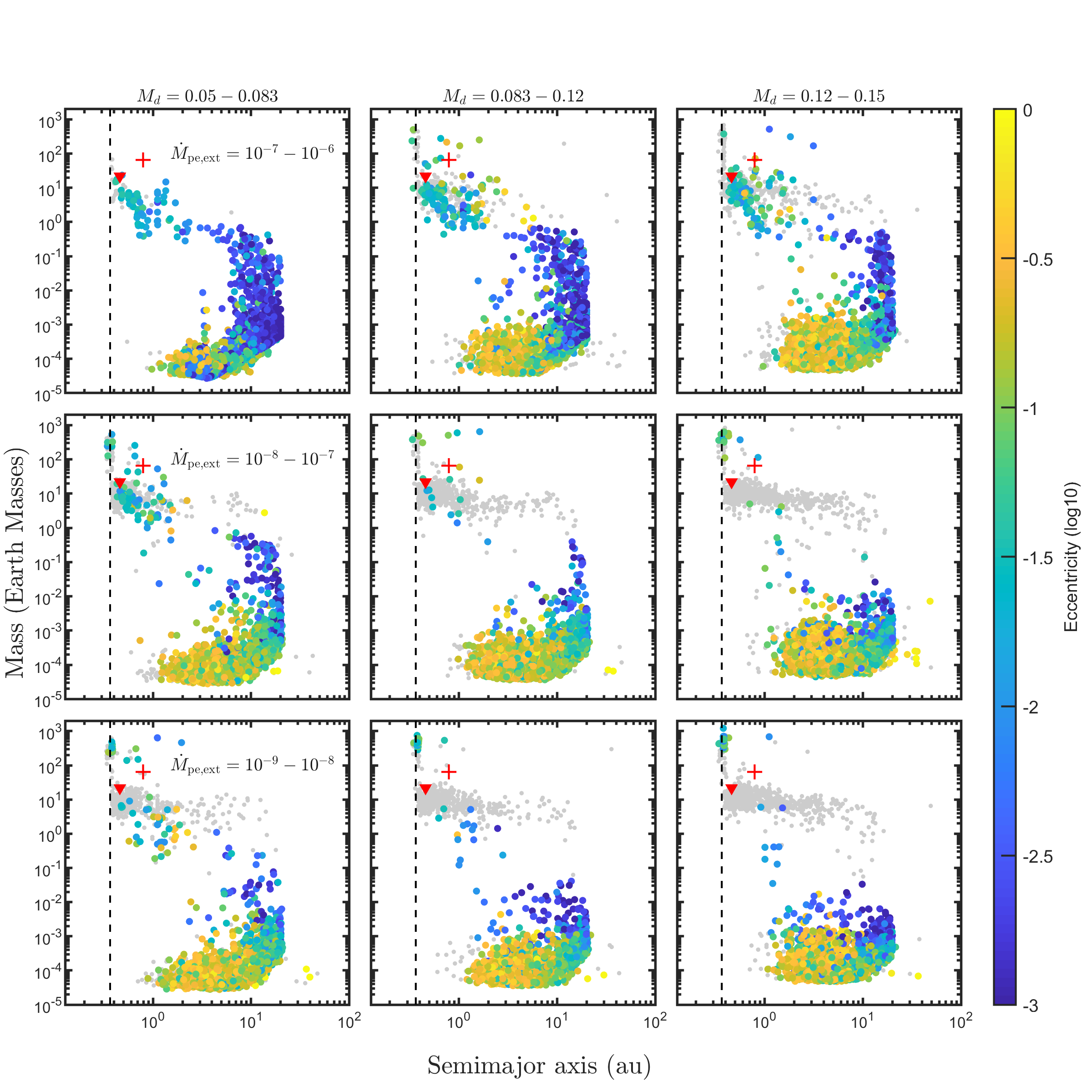}
\caption{Same as fig. \ref{fig:mva_large_4} but for viscosity $\alpha$ values between $10^{-3}$--$10^{-2.5}$.}
\label{fig:mva_large_3}
\end{figure*}

Whilst fig. \ref{fig:mva_large_4} shows the different mass versus semimajor axis for planets forming in low $\alpha$ discs, fig. \ref{fig:mva_large_3} shows the same results, but for planets forming in high $\alpha$ discs ($\alpha \ge 10^{-3}$).
As was shown in fig. \ref{fig:mva_alpha}, there is a difference in the distributions of planets that form in high $\alpha$ discs compared to low $\alpha$ discs, with more massive planets forming in higher $\alpha$ discs.
This is also evident when looking at the bottom two rows of fig. \ref{fig:mva_large_3} where giant planets are found to reside around the cavity region and the zone of dynamical instability.
Whilst most of the features seen above in fig. \ref{fig:mva_large_4} are equally seen in fig. \ref{fig:mva_large_3} (i.e. the effects of increasing disc mass or decreasing the external photoevaporation rate), there is also an interesting feature that arises in the smaller mass discs in high radiation environments.
Looking at the top left panel of fig. \ref{fig:mva_large_3}, it is interesting to see that no planets with masses $\sim m_{\rm p}>20\me$ were able to form.
This is due to planets not being able to accrete pebbles or gas for significant periods of times, since the strong external photoevaporation had effectively truncated the disc early into the disc lifetime.
The fast truncation of the discs resulted in the pebble production front reaching the disc outer edge early in the lifetime, which halted the mass flux of pebbles drifting into the inner system, such that they were not available to be accreted by planets \citep{Qiao23}.
Another effect limiting pebble accretion rates is the higher $\alpha$ values in these discs, which acted to stir up pebbles vertically, extending the pebble scale height.
With a larger pebble scale height, this reduced the efficiency of pebble accretion. 
These effects resulted in stunted growth for giant planet cores, leaving them little time to undergo runaway gas accretion and become giant planets.

Whilst the top left panel showed the effects on the population for low mass discs, the top right panel shows the population for high mass discs in high radiation intensity environments.
With more massive discs, their lifetimes are longer and there is more solid material to be accreted.
This resulted in a large number of terrestrials and super-Earths being able to form and migrate to the central cavity, with some being able to undergo runaway gas accretion and become giant planets.
In comparison with the same panel in fig, \ref{fig:mva_large_4}, a larger number of giant planets were able to form, however they were ejected from the system, as shown by the grey points seen near the dashed line that denotes the edge of the zone of dynamical instability.
The increase in ejections was due to the increased strength of the N-body interactions between multiple giant planets/giant planet cores, allowing planets to interact more effectively with the central binary, and therefore being ejected.

Similar to fig. \ref{fig:mva_large_4} the populations that formed in weaker radiation fields, shown by the bottom panels of fig. \ref{fig:mva_large_3}, contained more massive planets as they could form more easily in longer lived discs.
In the less massive discs (bottom left panel), a few giant planets formed, as well as a large number of super-Earth and Neptune mass planets.
Moving to the most massive discs (bottom right panel), a large number of giant planets were able to form, since the increase in disc mass allowed multiple planetary cores to accrete significant amounts of pebbles, which were then able to undergo runaway gas accretion and accrete large gaseous envelopes.
However with large numbers of giant planets forming, this again resulted in an increase in the number of ejected planets through mutual gravitational interactions and interactions with one of the binary stars.
These lost planets can be seen by the large number of grey points in the bottom right panel of fig.~\ref{fig:mva_large_3} for planets last seen orbiting within 1$\au$ and with masses between 1--100$\me$.
Comparing the bottom panels of fig. \ref{fig:mva_large_3} to those in fig.~\ref{fig:mva_large_4}, it is clear the semimajor axis distribution of the planets is slightly different, with those more massive planets forming in the higher $\alpha$ discs being found closer to the binary stars, since they were able to migrate there before opening gaps in the disc.
There was also limited outward migration in these discs, since the disc interior to a gap opening planet was able to accrete on to the central stars before the outer disc dispersed, unlike the scenario found for lower $\alpha$ discs.

\begin{figure}
\centering
\includegraphics[scale=0.6]{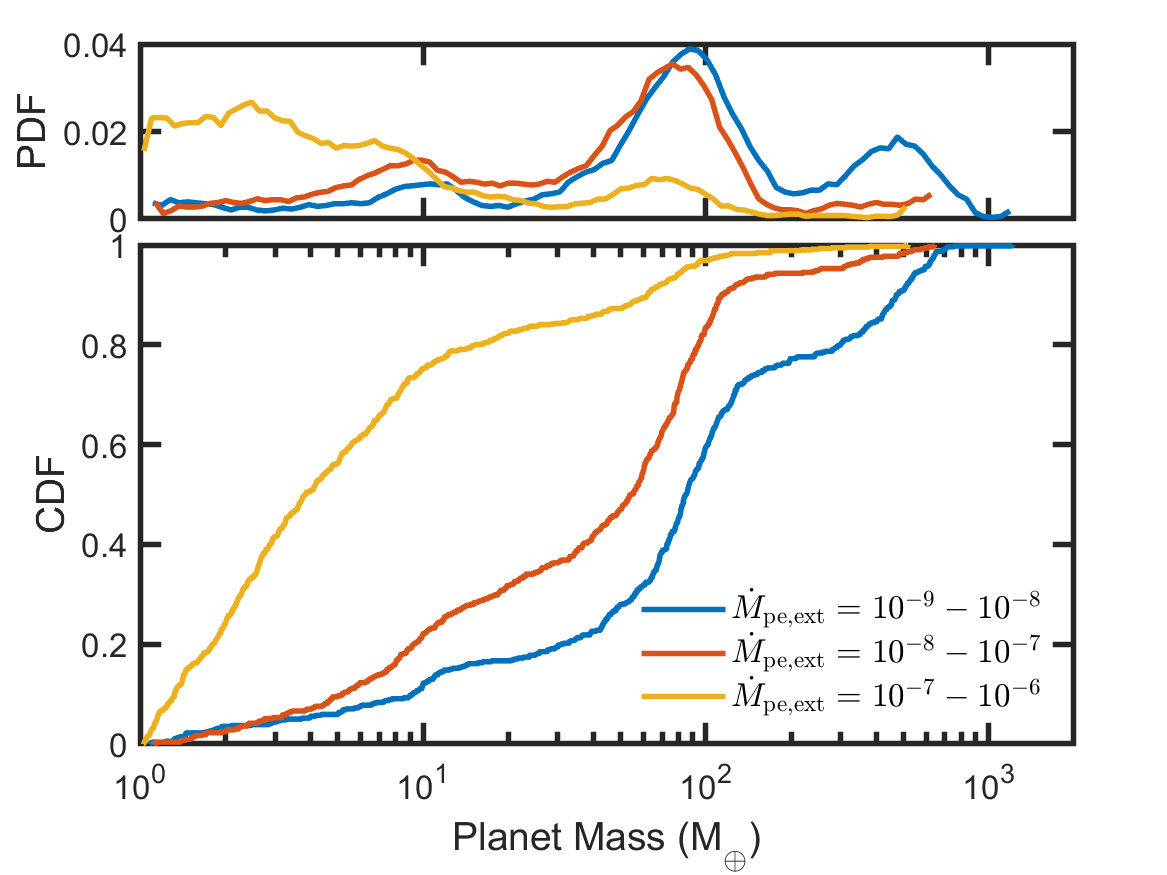}
\caption{Cumulative distribution functions (bottom panel) and probability distribution functions (top panel) of planet masses for those forming in discs with different ranges of the external photoevaporation rate. We show discs with mass loss rates between $10^{-9}$--$10^{-8}\msunyr$ (blue line), $10^{-8}$--$10^{-7}\msunyr$ (red line) and $10^{-7}$--$10^{-6}\msunyr$ (yellow line). We only calculate the distributions for planets with masses greater than 1$\me$.}
\label{fig:planet_mass_dist_ext}
\end{figure}

\begin{figure}
\centering
\includegraphics[scale=0.6]{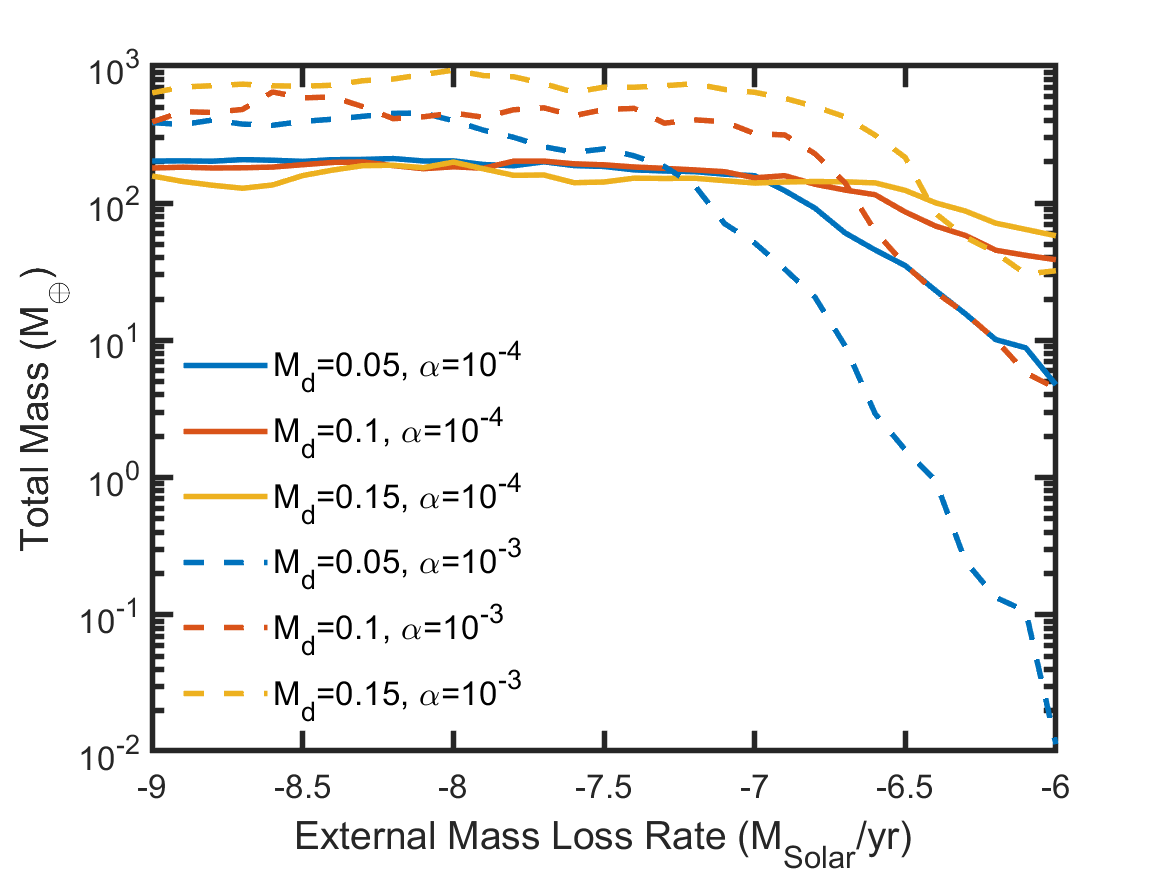}
\caption{The average total mass of simulated systems as a function of the external photoevaporation rate. Solid lines show those systems with low $\alpha$ viscosity discs, whilst dashed lines show high $\alpha$ viscosity discs. The initial disc mass is indicated by the colours: blue is equal to 0.05 $M_{\rm bin}$, red is equal to 0.1 $M_{\rm bin}$, and yellow is equal to 0.15 $M_{\rm bin}$.}
\label{fig:ext_dist_alpha}
\end{figure}

\subsection{Importance of the radiation environment}

The effects of external photoevaporation on the growth and migration of forming planets has only recently been examined.
\citet{Winter22} found that the gas accretion and migration of wide-orbit giant planets in protoplanetary discs can be suppressed by FUV-induced external photoevaporation.
More recently \citet{Qiao23} explored the effects of external photoevaporation on planetary cores accreting pebbles, finding that it can significantly hinder their growth.
As discussed above, we observe similar effects here, especially in more viscous discs, and those of lower mass.

Similar to fig. \ref{fig:planet_mass_dist}, in fig. \ref{fig:planet_mass_dist_ext} we show the cumulative distribution function (bottom panel) and the probability density function of planet masses as a function of the external photoevaporative mass loss rate. The blue lines represent small external mass loss rates, with the yellow lines showing large mass loss rates, and red lines showing intermediate mass loss rates. The effects of the external environment on planet mass is clear when comparing the blue lines to the yellow lines. Very few giant planets are able to form in strong external radiation environments as seen by the lack of planets with masses $m_{\rm p}>100 \me$. Even for planets with substantial gaseous envelopes, only $\sim20\%$ of planets have masses $m_{\rm p}>10 \me$. This is in contrast to the weaker radiation environments where the blue lines show that $\sim85\%$ of planets have masses $m_{\rm p}>10 \me$, and $\sim40\%$ of planets are giant planets. Looking at the probability distributions, there are interesting differences there, with strong radiation environments favouring planets of much lower masses and a small peak at $m_{\rm p}\sim70\me$, whilst the intermediate radiation environments show a large peak at $m_{\rm p}\sim70$--$80\me$, and weak radiation environments showing a double peak, with one at $m_{\rm p}\sim80\me$ and the other at $m_{\rm p}\sim500\me$. This shows the change in planet distributions that form in discs evolving under different external photoevaporation rates.

In fig.~\ref{fig:ext_dist_alpha} we show the average total mass of planets that form (i.e. the combined mass of all planets including those ejected) for circumbinary discs with varying external photoevaporation rates.
The solid lines show the total masses for circumbinary discs with low $\alpha$ values, whilst dashed lines show their more viscous counterparts. We differentiate the effect of initial disc mass with the different colours, with blue being low mass discs and yellow being high mass discs.
When looking at the left-hand side of fig.~\ref{fig:ext_dist_alpha} it is clear that the local radiation environment only has an effect once the initial external mass loss rate reaches $10^{-7.5} \msunyr$, in agreement with previous works that find that weak and/or shielded environments have limited effects on planet formation \citep{Winter22,Qiao23}.
The effect of higher $\alpha$ discs where giant planets are able to form can also be easily seen here, where the total planet exceeds $2 M_{\rm J}$ for more massive discs, whilst for low $\alpha$ discs, they generally converge to between 100--200 $\me$.
As the external photoevaporation rate increases, the drop in total planet mass is clear, especially in more viscous discs where in more extreme radiation environments ($\dot{M}_{\rm ext}\geq 10^{-6.5}\msunyr$), no planets with masses $m_{\rm p}>100\me$ are able to form.
Interestingly, the drop off in total planet mass is not as extreme in less viscous discs, due to the reduced amount of turbulence allowing for larger pebble accretion rates, resulting in some super-Earth and Neptune mass planets being able to form before the discs disperse.

\begin{figure*}
\centering
\includegraphics[scale=0.5]{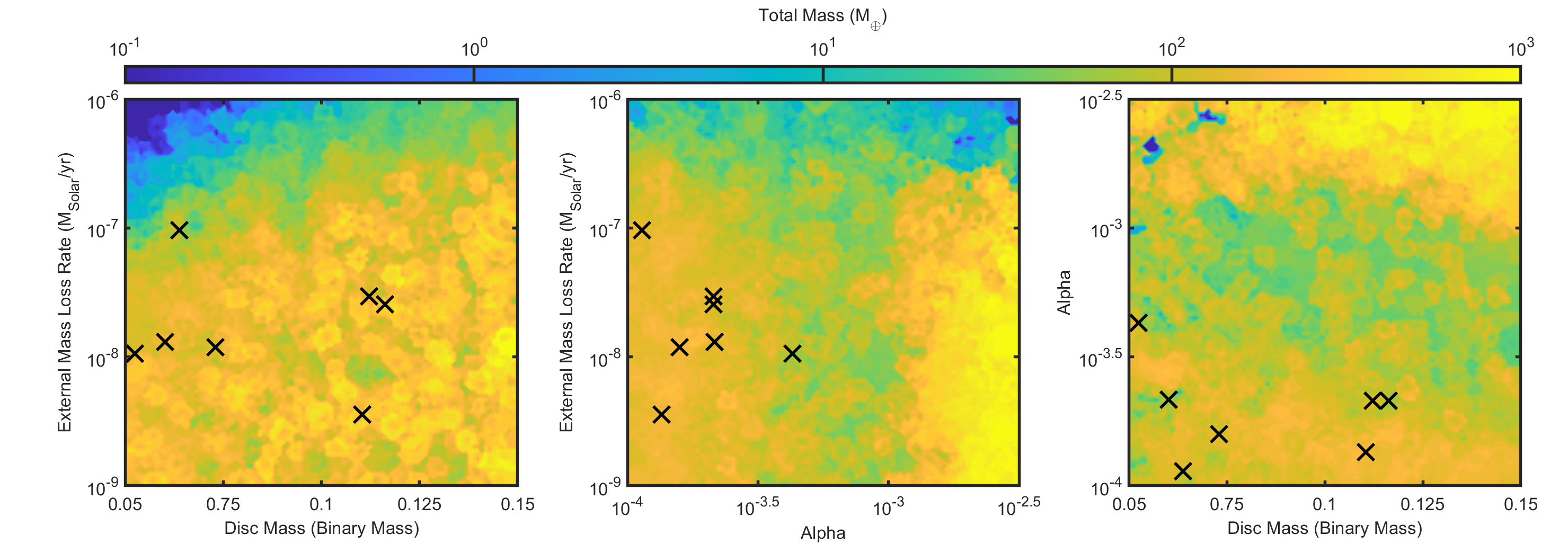}\\
\caption{Contour plots showing the total mass of all formed planets in systems as a function of different parameters. The left-hand panel compares disc mass to the external mass loss rate, the middle panel compares the viscosity parameter $\alpha$ to the external mass loss rate, and the right-hand panel compares the disc mass to $\alpha$. Black crosses denote the systems that best match TOI-1338/BEBOP-1.}
\label{fig:contour_tm}
\end{figure*}

\begin{figure*}
\centering
\includegraphics[scale=0.5]{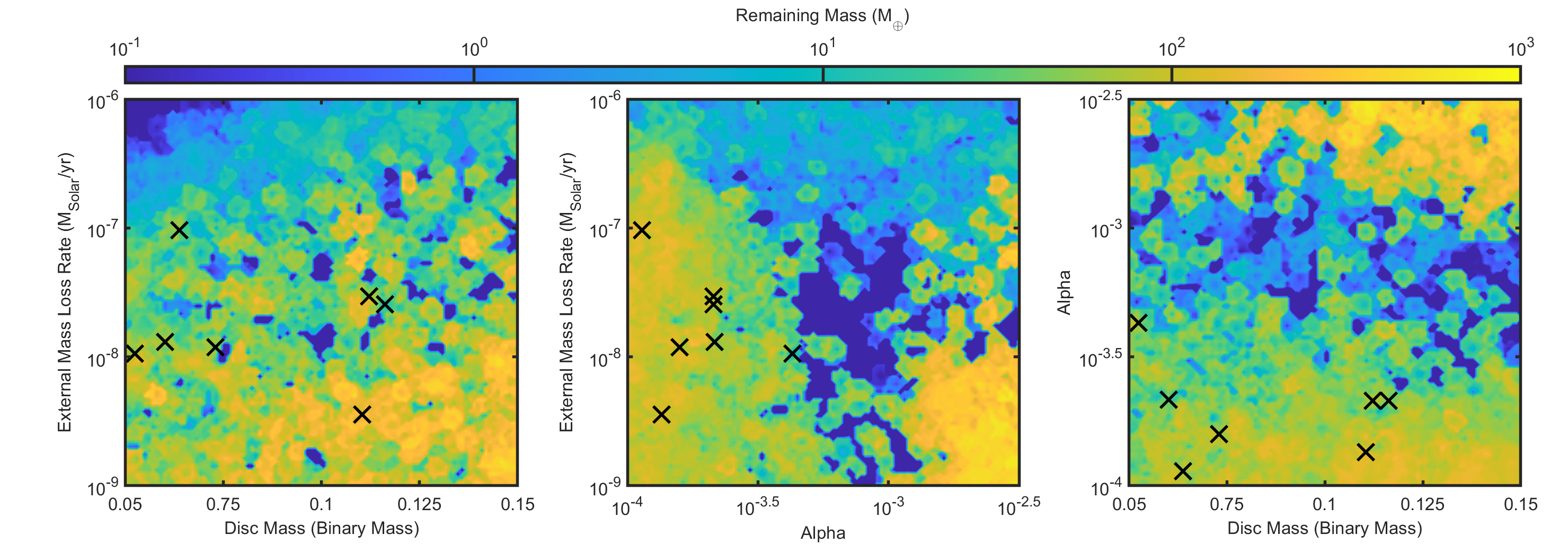}\\
\caption{Same as fig. \ref{fig:contour_tm} but the colour scale now showing the remaining mass for each simulation.}
\label{fig:contour_rm}
\end{figure*}

Overall, fig. \ref{fig:ext_dist_alpha} shows that external photoevaporation can have a large effect on the resulting planets and planetary systems.
As in previous works \citep[e.g.][]{Coleman22}, a mass loss rate of $10^{-8}\msunyr$ corresponds to a $G_0=300$ radiation environment\footnote{$G_0$ is taken as the flux integral over 912--2400\AA, normalised to the value in the solar neighbourhood \citep{Habing68}.}, and these results imply that in most observed star forming regions, external photoevaporation will have significant effects on the types and locations of planets that form.
This includes both star forming regions with strong UV fields  such as Orion, where the FUV field strengths range from $\sim30$--$\sim3\times 10^4$, and also weaker environments such as Taurus or Lupus where the FUV field strengths are approximately an 1--2 orders of magnitude lower on average than in Orion \citep{Winter18}.

\subsection{Complications due to {\textit{N}}-body interactions}

The previous sections discussed the effects of the initial disc mass, the local radiation environment and other disc properties.
We now show the effects that \textit{N}-body interactions have on the final planetary systems.
In fig. \ref{fig:contour_tm} the total mass of planets that formed in each system is represented by the colour in filled contour plots as a function of the different varying initial parameters.
In the left-hand panel we compare the external mass loss rate to the initial disc mass.
The middle panel compares the external mass loss rate against the disc viscosity parameter $\alpha$, whilst the right-hand panel compares $\alpha$ against the initial disc mass.
The black crosses denote the most TOI-1338/BEBOP-1 like systems, that will be discussed in Sect.~\ref{sec:BEBOP1}.

Looking at the left-hand panel of fig. \ref{fig:contour_tm}, a correlation can be seen in the total planet mass as a function of external photoevaporation and initial disc mass.
More mass is converted into planets in more massive discs and weaker radiation environments, i.e. towards the bottom right of the panel.
Conversely, in less massive discs in strong radiation environments, very little mass is accreted on to planets.
This is mainly a result of the lifetime of the disc, with lifetimes increasing towards the bottom right of the panel.
A similar effect can be seen in the middle panel where discs have longer lifetimes as the plot moves from the top to the bottom left.
The main difference in the middle panel here, is that the most massive planetary systems appear in the bottom right of the panel, due to giant planet cores being able to undergo runaway gas accretion before they open a gap in the discs.
This can be seen when comparing the bottom rows of figs. \ref{fig:mva_large_4} and \ref{fig:mva_large_3}.
When examining only $\alpha$, the middle panel shows that values between $\alpha=10^{-3.5}$--$10^{-3}$ are the most conducive to forming massive planets.
The right-hand panel compares the total masses when comparing $\alpha$ and the initial disc mass, where the total planet mass mainly increases as disc mass increases, whilst the most massive planets are again found in the top right of the panel where $\alpha$ is largest, allowing giant planets to form.

With fig. \ref{fig:contour_tm} showing contour plots for each set of parameters, fig. \ref{fig:contour_rm} shows a very similar plot, but only accounting for surviving planets, i.e. not including those planets that have collided with the central stars or been ejected from the system.
Comparing fig. \ref{fig:contour_rm} to fig. \ref{fig:contour_tm}, it is clear that the most notable differences are in the regions where the total mass indicates that multiple super-Earth to Neptune mass planets were able to form (i.e. >$20\me$).
This can very easily be seen in the left-hand and middle panels of fig. \ref{fig:contour_rm} where patches of blue denoting negligible planet mass remaining in the systems have now sporadically appeared where the total planet mass was high.
This indicates a large proportion of the mass locked in planets has been lost from the system, and in these cases, mainly due to ejections.
With there being so many planets of considerable mass, this allowed interactions between planets to excite eccentricities, allowing some to collide, but for most to cause them to interact strongly with the central binary, resulting in their ejection from the systems.

Our previous work also found that circumbinary systems ejected large numbers of planets. In examining the formation of Kepler-16 and Kepler-34, \citet{Coleman22b} found that between 6--10 planets were ejected on average from those systems, with most planets having masses less than that of Neptune.
In this work, over a broader parameter space, and around a different system, we find similar results.
On average, each simulation here ejected between 3--7 planets with 88\% of those planets having masses $m_{\rm p}\le \me$.
Therefore, in agreement with the work of \citet{Coleman22b}, circumbinary systems are an effective birthplace for free floating planets.
Indeed, recent microlensing surveys of free floating planets predict the the frequency to be $f=21^{+23}_{13} \rm star^{-1}$, with the total mass of free floating planets per star to be equal to $80^{+73}_{-43}\me\rm star^{-1}$, indicating slightly higher numbers than arise from our planet formation models \citep{Sumi23}.
In future work we will explore broader populations of close binary stars, to examine quantitatively the properties of free floating planets arising from those systems, which can then be compared to current and future microlensing surveys (e.g. \citet{Sumi23}, Nancy Grace Roman Telescope \citep{Spergel15,Bennett18} or the Large Synoptic Survey Telescope \citep{LSST_2019}).

The chaotic nature of the N-body interactions when considering them from an initial parameter point of view, make it extremely difficult to match specific planetary systems, since similar initial parameters can yield systems that are equally stable or unstable, as indicated by the patchiness of the blue regions amongst the yellow in fig. \ref{fig:contour_rm}.
Interestingly the systems with most mass in planets, i.e. the bottom right of the middle panels, showed that considerable mass remained in the systems even after dynamical interactions occurred once the gas disc had dispersed.
Even though multiple giant planets formed in such systems, there was generally at least one giant planet that remained orbiting the binary.
This could indicate that systems observed with giant planets, e.g. Kepler-1647 \citep{Kostov16}, may have had interesting dynamical histories.

\subsection{What disc properties led to the TOI-1338/BEBOP-1 planetary system?}
\label{sec:BEBOP1}

With the previous section looking at the parameter space as a whole, we now focus on determining the disc properties that best match the TOI-1338/BEBOP-1 planetary system.
To find the simulated systems that best match the TOI-1338/BEBOP-1 system, we devise a short list of constraints to then assign a score $S$ to each simulated system as follows
\begin{equation}
    S = S_1 \times S_2 \times S_3 \times S_4 \times S_5 \times S_6 \times S_7.
\end{equation}
The above equation is evaluated for each surviving planetary pair in each simulation.
The constraints $S_1$ and $S_2$ evaluate the relative semimajor axes between the two simulated planets and those in the TOI-1338/BEBOP-1 system:

\begin{equation}
\begin{minipage}{0.23\textwidth}
$S_{1} = \left\{ \begin{array}{ll}
\dfrac{a_{\rm p1}}{a_{\rm b}} ,& a_{\rm p1} \ge a_{\rm b}\\
\\
\dfrac{a_{\rm b}}{a_{\rm p1}} ,& a_{\rm p1} < a_{\rm b}\\\end{array} \right.$\\
\end{minipage}
\begin{minipage}{0.23\textwidth}
$S_{2} = \left\{ \begin{array}{ll}
\dfrac{a_{\rm p2}}{a_{\rm c}} ,& a_{\rm p2} \ge a_{\rm c}\\
\\
\dfrac{a_{\rm c}}{a_{\rm p2}} ,& a_{\rm p2} < a_{\rm c}\\\end{array} \right.$
\end{minipage}
\end{equation}
whilst $S_3$ and $S_4$ evaluate the relative differences in planetary masses:
\begin{equation}
\begin{minipage}{0.23\textwidth}
$S_{3} = \left\{ \begin{array}{ll}
\dfrac{m_{\rm p1}}{m_{\rm b}} ,& m_{\rm p1} \ge m_{\rm b}\\
\\
\dfrac{m_{\rm b}}{m_{\rm p1}} ,& m_{\rm p1} < m_{\rm b}\\\end{array} \right.$\\
\end{minipage}
\begin{minipage}{0.23\textwidth}
$S_{4} = \left\{ \begin{array}{ll}
\dfrac{m_{\rm p2}}{m_{\rm c}} ,& m_{\rm p2} \ge m_{\rm c}\\
\\
\dfrac{m_{\rm c}}{m_{\rm p2}} ,& m_{\rm p2} < m_{\rm c}\\\end{array} \right.$
\end{minipage}
\end{equation}

Similar to the procedure above, the next constraint, $S_5$ evaluates the ratio in semimajor axes between the simulated pair and those in the TOI-1338/BEBOP-1 system, whilst $S_6$ evaluates the relative mass ratios.
The final constraint, $S_7$ compares the total mass of planets with orbital periods of less than 1 year to the total expected mass of the TOI-1338/BEBOP-1 system.
This final constraint takes into account the point that our simulations are only run for 10 Myr, and so further N-body interactions could occur resulting in planets more similar to those observed from a system with equal mass amongst multiple planets with orbital periods less than 1 year.
With this score $S$ now calculated for each system, the lower the score that each system attains, the more similar that simulated system is to TOI-1338/BEBOP-1.

\begin{table}
\centering
\begin{tabular}{l|c}
Constraint & Value\\
\hline
$a_{\rm b}\ (\au)$ & 0.4607\\
$a_{\rm c}\ (\au)$ & 0.794\\
$m_{\rm b}\ (\me)$ & 21.6\\
$m_{\rm c}\ (\me)$ & 65.1\\
$a_{\rm c}/a_{\rm b}$ & 1.7235\\
$m_{\rm c}/m_{\rm b}$ & 2.9885\\
$m_{\rm b}+m_{\rm c} (\me)$ & 86.9\\
\end{tabular}
\caption{Values for the parameters use to test the distance from a simulated system to that of TOI-1338/BEBOP-1.}
\label{tab:distance}
\end{table}

\subsubsection{Best parameters for TOI-1338/BEBOP-1}

In figs. \ref{fig:contour_tm} and \ref{fig:contour_rm}, the black crosses show the parameters of the best fit TOI-1338/BEBOP-1 simulations.
The main correlation that comes from figs. \ref{fig:contour_tm} and \ref{fig:contour_rm}, is found in the middle panels, where the best fit systems occupy the parameter space of low viscosity discs ($\alpha\le 10^{-3.5}$) and in weaker radiation environments ($\dot{M}_{\rm ext}\le 10^{-7}\msunyr$).
Interestingly the initial disc mass has little effect on forming TOI-1338/BEBOP-1 like systems, mainly due to the other parameters allowing discs of all masses to have lifetimes sufficient to form the planets observed.
When comparing the best systems to those immediately in their vicinity, it is clear that a number of the best fits occupy regions of blue and yellow, where the systems either have little mass in planets or significant amounts.
This again highlights the chaotic \textit{N}-body interactions that shape planetary systems.

Whilst these results suggest that TOI-1338/BEBOP-1 did not form in a strong UV environment, the possible constraints it puts on $\alpha$ are interesting.
Observational studies have also hinted at low values for $\alpha$ \citep{Rosotti23}, whilst numerical studies of migrating circumbinary planets also find that low $\alpha$ values halt planets migrating in circumbinary discs, close to their observed locations \citep{Penzlin21}.
The observational estimates for $\alpha$ are determined from measurements of the disc radius, disc mass and
mass accretion rate.
Observing the Lupus cluster, \citet{Ansdell18} obtained values for $\alpha$ ranging from $\sim10^{-4}$ to $\sim10^{-1}$ with a median of roughly $3\times 10^{-3}$.
However a more recent study by \citet{Trapman20} implied that these values may be overestimated by up to a factor 10 and therefore, a median $\alpha$ of $3\times 10^{-4}$--$3\times 10^{-3}$ is more appropriate to explain observed accretion rates.
More recent observations of the dust disc around Oph163131 have found that very low levels of turbulence are required to recreate such a razor-thin pancake-like disc \citep{Villenave22}.
With most observations hinting at low levels of turbulence \citep[$\alpha<10^{-3}$; see table 3 in][for a recent review]{Rosotti23}, the indication from the sections above that to form a TOI-1338/BEBOP-1 like system requires low $\alpha$ values is in concordance with observations.
Indeed, looking at figs. \ref{fig:contour_tm} and \ref{fig:contour_rm} where the black crosses show the most TOI-1338/BEBOP-1-like systems, they all reside at low $\alpha$ with most between 1--2$\times10^{-4}$, and the odd outlier at 4$\times10^{-4}$.
When examining the most TOI-1338/BEBOP-1-like systems in regards to the initial disc mass and external photoevaporation rate, the predicted properties are less clear, except for indicating that they shouldn't be on the extreme end of the parameters studied, i.e. not too massive discs, or in extreme UV environments.
But in being able to provide indications for $\alpha$, these simulated systems show that by using observed exoplanet systems and current planet formation models in attempting to recreate them, could add valuable insight into the properties of protoplanetary or circumbinary discs.

\begin{figure}
\centering
\includegraphics[scale=0.6]{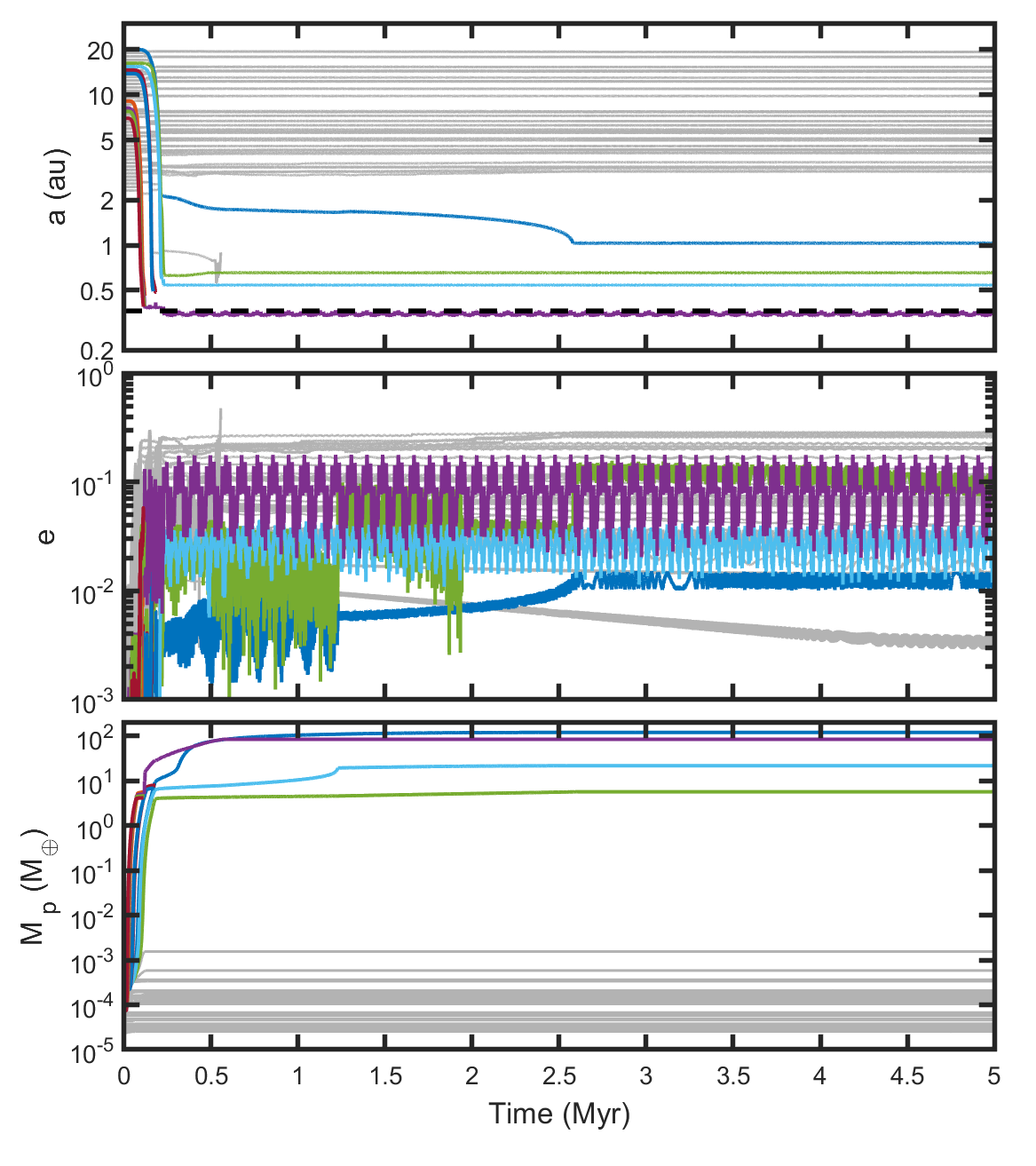}
\caption{Temporal evolution of planet semimajor axes (top), eccentricities (middle) and masses (bottom) for the most TOI-1338/BEBOP-1 like system described in sect. \ref{sec:BEBOP1}. The dashed horizontal black line denotes the outer edge of the zone of dynamical instability.}
\label{fig:sim_time}
\end{figure}

\subsubsection{TOI-1338/BEBOP-1 Example}

\begin{figure}
\centering
\includegraphics[scale=0.6]{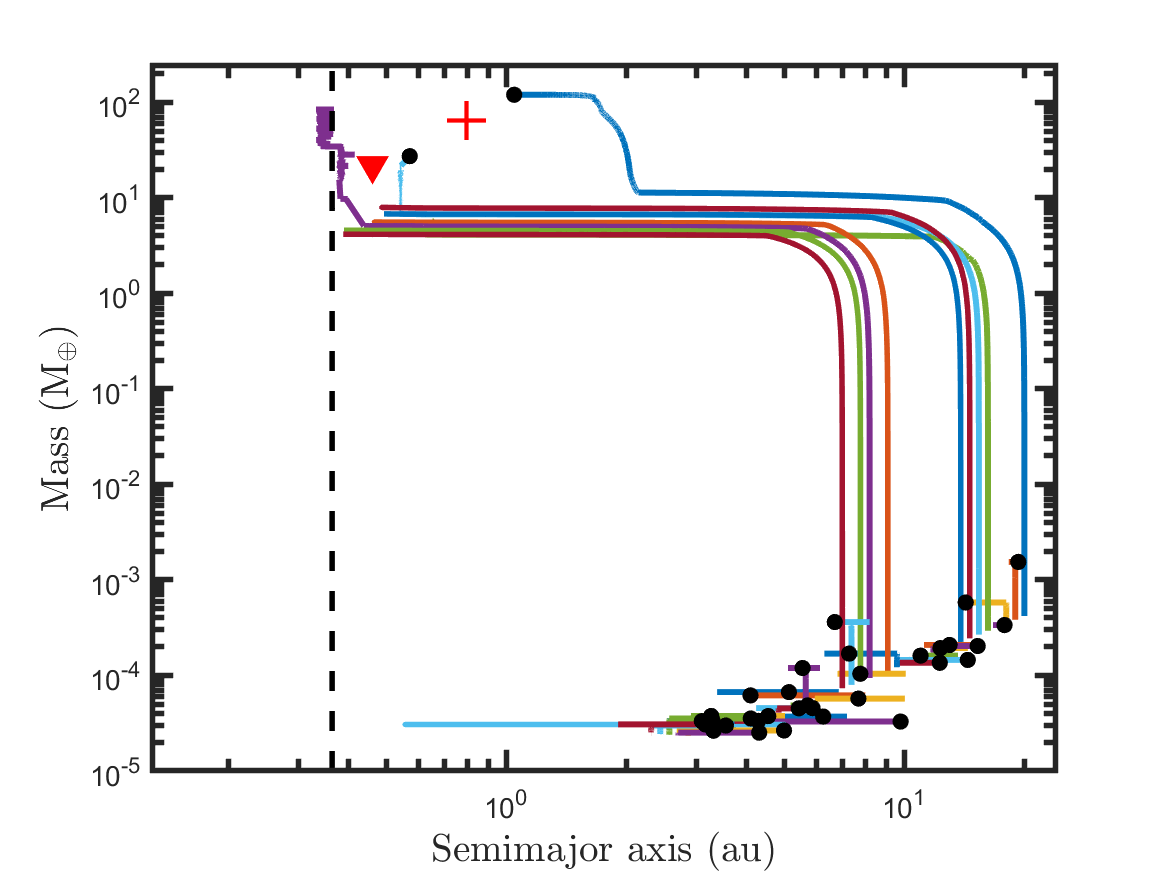}
\caption{Evolution of planet mass versus semimajor axis for the most TOI-1338/BEBOP-1 like system described in sect. \ref{sec:BEBOP1}. Filled black circles represent final masses and semimajor axes for surviving planets. Black dots represent the final masses and locations of the planets, whilst the red triangle and plus show the observed locations of TOI-1338b and BEBOP-1c respectively \citep{Kostov20,Standing23}. The dashed vertical black line denotes the outer edge of the zone of dynamical instability.}
\label{fig:sim_mva}
\end{figure}

We now briefly show and describe the most TOI-1338/BEBOP-1 like system that formed in our simulations.
Figure \ref{fig:sim_time} shows the temporal evolution of planet semimajor axes (top panel), eccentricities (middle panel) and masses (bottom panel) of one simulation that formed a system similar to TOI-1338/BEBOP-1. Grey lines show planets that had final masses $m_{\rm p} < 1\me$, with coloured lines representing planets with masses $m_{\rm p} \ge 1\me$.
Figure~\ref{fig:sim_mva} shows the mass versus semimajor axis evolution of forming planets, where the black points represent the final planet masses and semimajor axes.
The dashed black line in figs. \ref{fig:sim_time} and \ref{fig:sim_mva} shows the outer edge of the zone of dynamical instability for Kepler-16 \citep{Holman99}, while the red triangle and plus sign show the observed locations of TOI-1338b and BEBOP-1c respectively \citep{Kostov20,Standing23}.

The system formed as follows. As the pebble production front moved outwards, the planetary embryos on the most circular orbits were able to accrete significant quantities pebbles, allowing a number of them to grow to masses greater than an Earth mass. These planets began to migrate in towards the central cavity, and continued to accrete drifting pebbles as well as accreting gas.
With the viscosity in the disc being $\alpha =2.1\times10^{-4}$, planets were able to open gaps in the inner disc region when they reached masses $m_{\rm p}\sim10\me$.
This occurred for the planets represented by the purple and then the dark blue lines in fig. \ref{fig:sim_time} in the first 0.2 Myr.
With the innermost planet opening a gap in the disc and accreting gas, it slowed the migration of other planets stalling two planets, shown by the green and cyan lines in fig. \ref{fig:sim_time}.
Other planets that migrated inward were ejected from the system after interactions with the central binary stars.
After 0.5 Myr, the planets represented by the purple and blue lines reaches masses of $\sim 75\me$, and opened wide gaps in the disc where material was then unable to cross the gaps and accrete on to the planet as efficiently.
These planets then formed a resonant chain with two planets located interior to them, where they remained, slowly accreting until the end of the disc lifetime after 6.7 Myr.
After 7.6 Myr, a significant dynamical instability occurred causing the innermost giant planet (purple line) to be ejected from the system, whilst also destabilising the two less massive planets (green and cyan lines) causing them to collide and form a single planet.
This left two planets surviving in the inner system at the end of the simulation after 10 Myr with the system consisting of a 27 $\me$ planet at 0.57 $\au$ and a 120 $\me$ planet at 1.04$\au$.
Whilst these planets are slightly more massive and orbiting slightly further away than those in TOI-1338/BEBOP-1, they do represent the best fit system, and with slight variations in the initial parameters could lead to a much better fit when taking into the chaotic nature of {\it N}-body dynamics.

\section{Summary and conclusions}
\label{sec:conclusions}

In this work, we have explored a broad and comprehensive parameter space to investigate the possible formation pathway of the TOI-1338/BEBOP-1 circumbinary system.
We used an updated version of the {\textit{N}}-body code \textsc{mercury6} including the effects of a central binary, and coupled to this a self-consistent 1D viscously evolving disc model containing prescriptions for planet migration, accretion of gaseous envelopes, pebble accretion, disc removal through photoevaporative winds and prescriptions taking into account the effects of the central binary such as an eccentric cavity.
We explored the effects that the initial disc mass, the strength of the local radiation environment and the level of turbulence in the disc, have on the planetary systems that are able to form.
We focus on the recently discovered TOI-1338/BEBOP-1 to explore whether it can constrain parameters that led to its formation.
The main results from our study can be summarised as follows.

(1) The level of turbulence in the disc, indicated by the viscous $\alpha$ parameter, has a large effect on the types of planets and planetary systems that form, especially for giant planets. In low $\alpha$ discs, fewer giant planets form as their cores open wide gaps in the disc before that can undergo runaway gas accretion. This also allows some of these giant planets to migrate outwards, leading to a larger spread in the orbital radius distribution of such planets.
The mass and orbital radius distribution of super-Earth and Neptune mass planets is also subsequently affected by the level of turbulence in the discs.

(2) We find that the initial disc mass has only a small effect on the final outcomes. Increasing the disc mass leads to longer disc lifetimes, that allows slightly more massive planets to form. However with pebble accretion, the bulk of the formation of the planets occurs early in the disc lifetime, and so planets are typically formed before the consequences of the initial disc mass become apparent.

(3) Whilst the initial disc mass had little effect on the forming planetary systems, the external photoevaporation rate had a larger effect. When it was large, ($\ge10^{-7}\msunyr$) few giant planets were able to form, since the disc was very quickly truncated, resulting in little time for pebble accretion, leading to smaller cores that accreted gas much less efficiently. With smaller mass loss rates, more massive planets were able to form, and so the planetary systems appeared quite similar, independent of the external photoevaporation rate. Ultimately, the simulations showed that the external photoevaporation rate has a large impact on the total mass of forming planetary systems.

(4) Planetary systems such as TOI-1338/BEBOP-1 can point to possible parameters within planet formation scenarios. Here our planet formation models indicate that weak turbulence, $\alpha\le10^{-3.5}$, and weak UV environments produce systems that are most similar to TOI-1338/BEBOP-1. These similarities are in terms of the mass and location of the planets as well as the ratios between them. The results of this work also show the parameters where such systems are not able to form, or where N-body interactions become significant for the evolution of such systems.

(5) When the disc lifetimes are extremely long, and the discs are massive enough to form multiple giant planet cores, dynamical interactions between planets becomes important in deciding the outcomes of planetary systems. Here we find this to be the case by comparing the remaining planetary mass with the total mass and find that massive long lived discs can lead to a substantial decrease in the final planet mass in the system compared to that formed. This is due to multiple planets mutually driving up their eccentricities resulting in ejections from the system after interacting with one of the binary stars.
Interestingly there is a peculiar region of parameter space where this is less important, when $\alpha$ is large and the disc mass is massive, which is due to multiple giant planets forming, containing the bulk of the mass, and settling into stable orbital configurations.

The simulations we have presented here show that it is possible to use observed planetary systems to test planet formation models, and which indicate the parameters that lead to the formation of simulated analogues that closely resemble the real system. Examples of this include the strength of the local radiation environment through its effect on external photoevaporation, and the influence of turbulence in the disc, i.e. the magnitude of $\alpha$.
As shown above, the indication that $\alpha$ should be small is in agreement with expectations from observations of protoplanetary discs \citep[see ][ for a recent review]{Rosotti23}.
However, this was only a single system that has been tested in such a manner, and only when multiple systems are used to narrow down their possible formation scenario and the properties of their nascent protoplanetary discs, will we truly be able to apply stringent constraints on their formation.

In addition, planet formation models are continuously evolving and becoming more complex, adding in new physics.
For example, the models presented here assumed that the initial planetary embryos were already fully formed, however models are now being produced that include the formation of planetesimals and planetary embryos from an initial gas and dust disc \citep{Coleman21}.
In future work we will incorporate these model improvements and explore the differences that arise between them and simulations that start with fully formed planetary embryos.
Inclusion of a beginning-to-end model will allow other parameters to be tested to examine their importance in forming the resulting planetary systems, as well as allowing the planetary systems to possibly indicate the empirical ranges for those parameters.
It is only then that planet formation models will truly be able to be used to shed light on the exact formation pathway of specific planetary systems.

\section*{Data Availability}
The data underlying this article will be shared on reasonable request to the corresponding author.

\section*{Acknowledgements}
We thank the referee for useful comments on the paper, and John Chambers for providing an updated version of {\sc{mercury6}} including the integrators for circumbinary systems.
GALC was funded by the Leverhulme Trust through grant RPG-2018-418. RPN acknowledges support from STFC through grants ST/P000592/1 and ST/T000341/1.
MRS acknowledges support from the UK Science and Technology Facilities Council (ST/T000295/1), and the European Space Agency as an ESA Research Fellow.
This research utilised Queen Mary's Apocrita HPC facility, supported by QMUL Research-IT (http://doi.org/10.5281/zenodo.438045).
This work was performed using the Cambridge Service for Data Driven Discovery (CSD3), part of which is operated by the University of Cambridge Research Computing on behalf of the STFC DiRAC HPC Facility (www.dirac.ac.uk). The DiRAC component of CSD3 was funded by BEIS capital funding via STFC capital grants ST/P002307/1 and ST/R002452/1 and STFC operations grant ST/R00689X/1. DiRAC is part of the National e-Infrastructure.
This work was performed using the DiRAC Data Intensive service at Leicester, operated by the University of Leicester IT Services, which forms part of the STFC DiRAC HPC Facility (www.dirac.ac.uk). The equipment was funded by BEIS capital funding via STFC capital grants ST/K000373/1 and ST/R002363/1 and STFC DiRAC Operations grant ST/R001014/1. DiRAC is part of the National e-Infrastructure.
The authors would like to acknowledge the support provided by the GridPP Collaboration, in particular from the Queen Mary University of London Tier two centre.
This research received funding from the European Research Council (ERC) under the European Union's Horizon 2020 research and innovation programme (grant agreement n$^\circ$ 803193/BEBOP)

\bibliographystyle{mnras}
\bibliography{references}{}

\label{lastpage}
\end{document}